\documentclass[aps,amssymb,amsfonts,amsmath,a4paper,preprint]{revtex4}
\usepackage{longtable}
\newcommand{\mms}{m$^2$s$^{-1}$}

\newif\ifpdf
\ifx\pdfoutput\undefined
  \pdffalse
\else
  \pdfoutput-1
  \pdftrue
\fi
\ifpdf
     \usepackage[pdftex]{graphicx}
     \DeclareGraphicsExtensions{.pdf}
\else
     \usepackage[dvips]{graphicx}
     \DeclareGraphicsExtensions{.EPS}
\fi

\normalsize
\makeatletter
\renewcommand{\@seccntformat}[1]{\large \csname the#1\endcsname .
\hspace{0.5em}}
\makeatother
\numberwithin{equation}{section}

\bibpunct{(}{)}{;}{a}{}{,}

\begin{document}

\title{Evidence for a $k^{-5/3}$ spectrum from   
 the EOLE Lagrangian balloons in the low stratosphere   
}

\author{Guglielmo Lacorata$^1$, Erik Aurell$^2$, Bernard Legras$^3$ and 
Angelo Vulpiani$^4$}

\address{$^1$CNR, Institute for Atmospheric and Climate Sciences, 
Lecce, Italy
}
\address{$^2$ KTH Royal Institute of Technology, Department of Physics, 
Stockholm, Sweden}
\address{$^3$ Ecole Normale Sup\'erieure, Laboratoire de 
M\'et\'eorologie Dynamique, UMR8539, 
Paris, France
}
\address{$^4$ University of Rome ''La Sapienza'', Department of Physics, 
 and \\ INFM (UdR and SMC), Rome, Italy
}

\begin{abstract}
{\noindent 
The EOLE Experiment is revisited to study turbulent
processes in the lower stratosphere circulation from a Lagrangian
viewpoint and resolve a discrepancy on the slope of the atmospheric
energy spectrum between the work of Morel and Larchev\^eque (1974) and
recent studies using aircraft data.  Relative dispersion of
balloon pairs is studied by calculating the Finite Scale Lyapunov
Exponent, an exit time-based technique which is particularly efficient
in cases where processes with different spatial scales are interfering.
Our main result is to reconciliate the EOLE dataset with recent studies
supporting a $k^{-5/3}$ energy spectrum in the range 100-1000~km.  Our
results also show exponential separation at smaller scale, with 
characteristic time of order 1 day,  and agree
with the standard diffusion of about $10^7$~\mms at large scales.  A
still open question is the origin of a $k^{-5/3}$ spectrum in the
mesoscale range, between 100 and 1000 km.}
\end{abstract}

\maketitle

\newpage

\section{Introduction}

The EOLE project consisted in the release of 483 constant-volume
pressurized balloons, in the Southern Hemisphere mid-latitudes
throughout the period September 1971-March 1972, at approximately 200
hPa.  The life-time of these balloons was from a few days to about one
year, with an average value of about 120 days.  Their motion was
basically isopycnal except for small diurnal volume variations of the
envelop of less than 1\% and inertial oscillations of a few meters in
the vertical, excited by wind shear and small-scale turbulence.  The
position of the balloons and meteorological parameters were
periodically transmitted to satellite by ARGOS system.  The
trajectories of the EOLE experiment still provide nowadays the most
extensive data set of experimental quasi-Lagrangian tracers in the
atmosphere for observing the properties of the medium-to-large-scale
motion at the tropopause level.

Both Eulerian and Lagrangian analyses have been performed by several
authors.  Morel and Desbois (1974) deduced the mean circulation around
200 hPa from the balloon flights, as formed by a mid-latitude zonal
stream with a meridional profile characterized by a typical velocity
$\sim 30$ ${\rm ms^{-1}}$ inside the jet, overimposed to meridional
velocity field disturbances of much smaller intensity, $\sim 1$ ${\rm
ms^{-1}}$, and to residual standing waves acting as spatial
perturbations of the zonal velocity pattern, producing the typical
shape of a meandering jet.  These results have been largely confirmed
by operational analysis since then.  Morel and Larchev\^eque (1974),
hereafter ML,
investigated the synoptic-scale turbulent properties.  They measured
the mean square relative velocity and the relative diffusivity of
balloon pairs, and found essentially two major regimes for Lagrangian
dispersion: exponential separation for time delays less than 6 days and
standard diffusion for larger times.  These authors also observed that
the scaling of the relative diffusivity with the separation length
between two balloons agreed with a direct 2D turbulent cascade, with
energy spectrum $E(k) \sim k^{-3}$, or steeper, in the range
100-1000~km.  Further Eulerian analyses of large-scale velocity
spectra by Desbois (1975) were compatible with the scenario proposed by
Morel and Larchev\^eque (1974) about isotropic and homogeneous 2D
turbulence with a $k^{-3}$ energy distribution up to scales $\sim
1000$~km.

Later, other authors reached for different conclusions after observing
energy spectra in the low stratosphere, measured from experimental data
recorded from commercial aircraft flights, Gage (1979), Lilly (1983),
Nastrom and Gage (1985).  Their picture suggested a 2D turbulent
inverse cascade, characterized by the $E(k) \sim k^{-5/3}$ spectrum,
inside the interval of scales 100-1000 km.

More recently, Lindborg and Cho (2000, 2001a and 2001b) computed
velocity spectra using data recorded during the MOZAIC program and also
found a $k^{-5/3}$ spectrum.  They suggested a dynamical mechanism
different from 2D inverse cascade where energy is injected in the large
scales by breaking Rossby-gravity waves and generates a chain process
down to smaller scales.  Their hypothesis is supported by the
observation of a downscale energy flux, whereas 2D inverse cascade
should exhibit upscale energy flux (Lindborg and Cho, 2000).

We wanted to reconsider this issue by performing a new analysis of the
relative dispersion properties of the EOLE balloons within the
framework of dynamical system theory.  Relative dispersion properties
are analyzed through the computation of the Finite-Scale Lyapunov
Exponent, or shortly FSLE (Aurell et al.  1997, Artale et al.  1997,
Boffetta et al.  2000a).  The FSLE is based on the growth rate
statistics of the distance between trajectories at fixed scale, and is
a better tool at analyzing scale dependent properties than plain
dispersion, as explained below.  This new method has been already
exploited for studies of relative dispersion in atmospheric and oceanic
systems (Lacorata et al.  2001, Joseph and Legras 2002, Boffetta et al.
2001, LaCasce and Ohlmann 2003, Gioia et al.  2003) and also in
laboratory convection experiments (Boffetta et al.  2000b).

This paper is organized as follows: in section \ref{sec:fsle} we
describe the FSLE methodology; section \ref{sec:analysis} contains the
results obtained from our analysis of the EOLE experimental data; in
section \ref{sec:conclusions} we discuss the physical information that
can be extracted from this paper and possible perspectives.
       
\section{Finite-scale relative dispersion}
\label{sec:fsle}

Generally speaking, most flows exhibit a range of scales over which
fluid motion is expected to display different characteristics: a
small-scale range where the velocity variations can be considered as a
smooth function of space variables; a range of intermediate lengths
corresponding to the coherent structures (and/or spatial fluctuations)
present in the velocity field over which velocity variations are rough
but highly correlated; a large-scale range over which spatial
correlations have definitely decayed.  In each of these ranges,
relative dispersion between trajectories is governed by a different
physical mechanism (chaos, turbulence, diffusion) which can be, in
principle, identified from the observations.  In fully developed
three-dimensional turbulence, motion is only smooth under the
Kolmogorov dissipative scale.  In the free stratified atmosphere (above
the planetary boundary layer), turbulence is a relatively rare event:
motion is most often smooth but for some localized convective or
turbulent events, associated with mesoscale systems, that mix momentum
and tracers.  Hence one expects to find a smooth (chaotic) dispersion
range ended at a scale characteristic of the spacing of mixing events,
followed by a range covering the mesoscale to synoptic range, and
finally standard diffusion at planetary scale.  This view is supported
by the ubiquitous observation of long-lived laminated structures in the
free troposphere (Newell {\it et al.}, 1999).

In order to fix some terminology, we will use both symbols $R$ and
$\delta$ for indicating the distance between balloons: the former will
be considered as a quantity function of time, the latter as an
argument for scale-dependent functions.

\subsection{\normalfont\itshape Diffusion and Chaos}

Diffusion is characterized in terms of diffusion coefficients, 
related to the elements of a diffusion tensor defined as 
\begin{equation}
  D_{ij} = \lim_{t \to \infty} {1 \over 2t} 
  \langle (x_i(t)-\langle x_i(0) \rangle)(x_j(t)-\langle x_j(0) 
  \rangle )\rangle
  \label{eq:difftens}
\end{equation} 
where $x_i(t)$ and $x_j(t)$ are the $i$-th and $j$-th coordinates
at time $t$, with $i,j = {1,2,3}$.  The average operation $\langle
\rangle$ is meant to be performed over a large number of particles.
The diagonal elements are just the diffusion coefficients.  When the
$D_{ii}$'s are finite, then the diffusion is standard.  This means that
at long times, after the Lagrangian velocity correlations have decayed
(Taylor, 1921), the variance of the particle displacement follows the
law: 
\begin{equation}
  \langle || {\bf x}(t)-\langle {\bf x}(0)\rangle ||^2\rangle
  \simeq 2Dt
  \label{eq:brown}
\end{equation}

In presence of a velocity field characterized by coherent structures, 
it is more useful to observe the relative dispersion between the trajectories, 
rather than the absolute dispersion from the initial positions, given by
(\ref{eq:brown}), which is unavoidably dominated by the mean advection. 

In the case of the EOLE experiment, where observing the expansion of
balloon clusters with more than two elements is a rare event (ML) a
measure of relative dispersion is given by the mean square
inter-particle distance:
\begin{equation} 
  \langle R^2(t) \rangle =
  \langle ||{\bf x}^{(m)}(t)-{\bf x}^{(n)}(t)||^2 \rangle
  \label{eq:reldisp}
\end{equation} 
averaged over all the pairs $({\bf x}^{(m)}, {\bf x}^{(n)})$,
where $m$ and $n$ label all the available $N$ trajectories.  Notice
that the norm in (\ref{eq:brown}) and (\ref{eq:reldisp}) must be
defined accordingly to the geometry of the fluid domain, i.e. in the
atmosphere we use the arc distance on the great circle of the Earth
connecting the two points.  The quantity in (\ref{eq:reldisp}) can be
measured for both initially close pairs, balloons released from the
same place at short time delay, and so-called {\it chance pairs},
balloons initially distant which come close to each other at a certain
time and, later, spread away again (ML).
Consistency of the average in eq.  (\ref{eq:reldisp}) requires all the
trajectory pairs to have nearly the same initial distance, a 
condition which strongly limits the statistics.  At long times, $\langle
R(t)^2 \rangle$ defined in eq.  (\ref{eq:reldisp}) is expected to
approach the function $4 D t$, where the $4$ factor accounts for
relative diffusion.  When it happens that $\langle R(t)^2 \rangle \sim
t^{2\nu}$ with $\nu > 1/2$, instead, the Lagrangian dispersion is
considered as super-diffusion.  A well-known example is the
Richardson's law for the particle pair separation in 3D turbulence, for
which $\nu=3/2$ (Richardson 1926; Monin and Yanglom 1975).
   
On the other hand, in the limit of infinitesimal trajectory
perturbations, much smaller than the characteristic lengths of the
system, the evolution of the particle pair separation is characterized
by the Lyapunov exponent (Lichtenberg and Lieberman, 1982), such that
\begin{equation}
  \lambda = \lim_{t \to \infty} \lim_{R(0) \to 0} 
  {1 \over t} \ln {R(t) \over R(0)}
  \label{eq:LE}
\end{equation} 
If $\lambda > 0$ the growth is exponential and the motion is said
chaotic.  Chaos is a familiar manifestation of non linear dynamics,
leading to strong stirring of trajectories (Ottino, 1989).  The
process, for example, of repeated filamention around the polar vortex
is basically due to Hamiltonian chaos (Legras and Dritschel, 1993).
For finite perturbations within a smooth flow, the properties of
exponential separation are observed for a finite time.

\subsection{\normalfont\itshape Finite-Scale Lyapunov Exponent}

The idea of FSLE (Aurell et al.  1997; Artale et al.  1997), was
formerly introduced in the framework of the dynamical systems theory,
in order to characterize the growth of non-infinitesimal perturbations
(i.e. the distance between trajectories).  If $\delta$ is the scale of
the perturbation, and $\langle \tau(\delta) \rangle$ is the mean time
that $\delta$ takes to grow a factor $r > 1$, then the FSLE is defined
as 
\begin{equation}
  \lambda(\delta) = {1 \over \langle \tau(\delta) \rangle} \ln r
  \label{eq:fsle}
\end{equation} 
The average operation is assumed to be performed over a large ensemble
of realizations.  For factors $r$ not much larger than $1$,
$\lambda(\delta)$ does not depend sensitively on $r$.  If $r=2$ then
$\langle \tau(\delta) \rangle$ is also called doubling time.
Operatively, $N+1$ scales are chosen to sample the spatial range of
perturbation, $\delta_0 < \delta_1 <...< \delta_N$, and a growth factor
$r$ is defined such that $\delta_{i} = r \cdot \delta_{i-1}$ for $i=1,N$.
Let $l_{min}$ and $l_{max}$ be the smallest and the largest
characteristic length of the system, respectively.  If $\delta_0 \ll
l_{min}$ then the FSLE characterizes the doubling time of infinitesimal
perturbation.  In the opposite side of the range, if $\delta_N \gg
l_{max}$ then the FSLE follows the scaling law of diffusion
$\lambda(\delta) \sim \delta^{-2}$ for $\delta \to \delta_N$, as can be
deduced by noticing that the mean square particle distance must grow
linearly in time, see (\ref{eq:brown}).  In general, if the mean square
size of a tracer concentration follows the $\langle R^2 \rangle \sim
t^{2\nu}$ law, the FSLE scales as $\lambda(\delta) \sim
\delta^{-1/\nu}$.  As we have seen before, for standard diffusion
$\nu=1/2$ while for Richardson's super-diffusion $\nu=3/2$.  The main
interest of FSLE is to select processes occurring at a fixed scale.  We
stress that definition (\ref{eq:fsle}) differs substantially from
\begin{equation}
  \lambda'(\delta) = {1 \over \langle R^2 \rangle} 
  {d \langle R^2 \rangle \over dt}|_{\langle R^2 \rangle =\delta^2} 
  \label{eq:lambdaprime}
\end{equation}
defined in terms of the mean square relative displacement, because of
the different averaging procedures in the two cases: $\langle R^2
\rangle$ is computed at fixed time while $\tau(\delta)$ is computed at
fixed scale.  As a result, a physical situation which is well
characterized in terms of FSLE, either for scaling properties or the
existence of transport barriers, may be less easily characterized by
studying the time growth of trajectory separation (Boffetta et
al., 2000b; Joseph and Legras, 2002).  One reason is that $\langle
R^2(t) \rangle$ depends on contribution from different regimes, as
seen, for example in $3D$ turbulence where a dramatic dependence of
$R^2(t)$ upon $R^2(0)$ is observed, even at very large Reynolds number
(Fung {\it et al.}, 1992)
   
In cases where advection is strongly anisotropic, e.g. in presence of a
structure like the stratospheric jet stream, it may be useful to define
the FSLE in terms of meridional (cross-stream) displacement only: 
\begin{equation}
  \lambda_{mer}(\delta^{(mer)}) = {1 \over 
  \langle \tau(\delta^{(mer)}) \rangle } \ln r
  \label{eq:fslemer}
\end{equation}
where $\delta^{(mer)}$ is the latitude distance (or meridian arc) 
between two points. 
  
Informations about the 
relative dispersion properties are also extracted by another 
fixed-scale statistics, 
the Finite-Scale Relative Velocity (FSRV), 
named by analogy with FSLE, that is defined as  
\begin{equation}
  \nu_2(\delta)  =  \langle \delta {\bf v}(\delta)^2 \rangle
  \label{eq:velgrad}
\end{equation}
where  
\begin{equation}
 \delta {\bf v}(\delta)^2  = ({\dot{\bf x}^{(1)}} -{\dot{\bf x}^{(2)}})^2  
\label{eq:veldiff}
\end{equation}
is the square  
Lagrangian velocity difference between two trajectories, ${\bf x}^{(1)}$ 
and ${\bf x}^{(2)}$,  
on scale $\delta$, that is  for $|{\bf x}^{(1)}-{\bf x}^{(2)}|=\delta$.
The FSRV can be regarded as the 2nd order 
structure function of the Lagrangian velocity difference and provides 
a complementary analysis to the FSLE diagnostics. In particular, in the regime  
of Richardson's super-diffusion, the expected behavior 
for the FSRV is $\nu_2(\delta) \sim \delta^{2/3}$. 

We report  in the next section the results of our analysis. 
 
\section{Analysis of the EOLE Lagrangian data}
\label{sec:analysis}

After a preliminary data check, the number of balloons selected 
for the analysis has been reduced to 382. This has been obtained by discarding
ambiguous ident numbers (some ident numbers have been used twice during the 
campaign), discarding trajectories that cross the equator and short
tracks of less than 10 points. 

Successive points along a balloon trajectory were mostly recorded at a
time interval of $10^{-1}$ day ($2.4$ hours), but the overall
distribution of the raw data does not cover uniformly the time axis.
Hence, each of the coordinates (longitude and latitude) of every
balloon trajectory has been interpolated in time by a cubic spline
scheme, with a sampling rate of 25 points per day.  Because of possible
data impurities, each Lagrangian velocity value is monitored at every
time step ($0.04$ day) and data segment with abnormally fast motions
are discarded.

As pointed out by ML, a way to measure the
dispersion between balloons is waiting for one of them to get close to
another one, at a distance less than a threshold $\delta_0$, and then
observing the evolution of their relative distance in time.  This
procedure is repeated for each balloon trajectory until the whole set
of pairs is analyzed.  The dataset includes original pairs of balloons that
were launched within a short time interval and chance pairs of 
balloons meeting suddenly after a number of days. For the largest values
of the threshold used in this study, the number of chance pairs largely exceeds 
the number of original pairs. In this way, global properties of the Lagrangian
transport are extracted from the contributions of balloon pairs
randomly distributed all over the domain. The number of balloon pairs 
and its evolution as the separation crosses the $N$ scales defined above
is described by Table 1.

In Figure \ref{fig:eoledisp100} four global relative
dispersion curves are plotted, referring to four different initial 
thresholds $\delta_0=$ 25 km, 50 km, 100 km and 200 km. The statistical 
samples vary roughly in proportional way with $\delta_0$.    
Relative dispersion depends sensitively, as expected, on the initial 
conditions; the four curves meet together for separation larger 
than about 2500 km and saturation begins for separation larger than 4500 km, leaving 
room only for a short standard diffusive regime between these two separations and
over a time duration of less than 10 days.
The eddy diffusion coefficient,
$D_E$, estimated by fitting the linear law $4D_Et$, results in
$D_E \simeq 2.9 \; 10^6$~\mms, a value compatible with what was found by ML.
The pre-diffusive regime is not very clear, we can say that the behavior 
of the balloon separation looks like a power law with exponent (changing in time) 
between 3 and 1.  

We report in Figure
\ref{fig:lyap} the mean logarithmic growth of the balloon relative 
separation over all pairs selected by 
the threshold 25 km. 
At very short times ($<$ 1 day) the slope 
corresponds to an exponential growth rate with $e$-folding time $\simeq$ 0.4~day 
that we consider 
as a rough estimate of the LLE.   
At later times, the slope gradually decreases as the separation growth tends to  
a power law regime. 
 In the same figure we also show the mean logarithmic growth 
of the inter-balloon distance computed for 
two 4-element clusters (that we label as 'A' and 'B'), 
launched with a time interval of 3 days between them.  
 A linear behavior (exponential growth) for both clusters is observed for short
intervals; we observe as 
 the exponential regime lasts longer for the 'A' cluster 
($\simeq 3$ days) than for the 'B' cluster ($\simeq 1$ days).  
This illustrates
the fact that the duration of a dispersion regime, here the chaotic
 one, may exhibit large fluctuations generally due to 
 different meteorological conditions.  As a result, average
time-dependent quantities, like $\langle R^2(t) \rangle$, sample
different regimes at once and are poor
diagnostics of dispersion properties. 
Incidentally, in ML the behavior of the relative dispersion between 
100 and 1000 km is fitted by means of an exponential with characteristic 
e-folding time $\simeq 2.7$ days (see ML, figure 8), which is compatible 
with the growth rate of Figure \ref{fig:lyap} between the two horizontal lines 
marking the range 100-1000 km, 
if one wants to fit it with an exponential curve branch.  

Figure \ref{fig:eolefsle3} shows the global FSLE relatively to the same four  
initial thresholds used for the relative dispersion and setting the
amplification ratio $r$ to $\sqrt{2}$.  The main result of
this study is that up to about 1000~km there is evidence of
Richardson's super-diffusion, compatible with a $k^{-5/3}$ spectrum,
displayed by the behavior $\lambda(\delta) = \alpha \delta^{-2/3}$. 
The best fit is obtained for the initial thresholds 100 and 200 km which
encompass a much larger number of pairs than smaller thresholds (see Table 1).
The quantity $\alpha^3$ is physically related to the mean relative kinetic energy 
growth rate (for unit mass) between balloons moving apart. 
Standard diffusion is approached at scales larger than 2000~km. 
The value of the eddy diffusion coefficient is 
 estimated by fitting the FSLE in the 
diffusive range with $(4 \ln r) D_E \delta^{-2}$, as shown 
in Boffetta {\it et al.} (2000a) by means of a dimensional argument. 
 We find that this value is $D_E \simeq 10^7$ \mms.  
Notice
that the initial threshold does not affect very much the
general behavior, except for obvious changes in the statistical 
samples. 

Figure \ref{fig:eolefsle100} shows global (mainly zonal) and meridional
($\lambda_{mer}$, see (\ref{eq:fslemer})) FSLE of the balloon pairs with
initial threshold $100$~km.  We find that the dispersion is basically
isotropic up to scales of about 500~km, which is in rough agreement
with the results of Morel and Larchev\^eque (they give a value three
times larger but their analysis, see their Fig.  7, does not display a
well-defined cut-off).  At scales larger than 500~km, the two
components of the FSLE decouple and the meridional dispersion rate
follows the standard diffusion law $\sim \delta^{-2}$ with a meridional
eddy diffusion coefficient $D_E \sim 10^6$~\mms.  

In order to compute 
the FSRV, the relative velocity between balloons is approximated 
by the finite difference formula $(|R(t+\Delta t)| - |R(t)|)/\Delta t$, 
where $|R(t)|$ is the absolute value of the great circle arc between two 
balloons at time $t$ and $\Delta t=0.04$ day 
is the time interval between two successive 
points along a trajectory.  
The properties of the Lagrangian relative velocity 
are shown in Figure \ref{fig:rv}.  The FSRV 
confirms the results obtained with the FSLE: 
between 100 and 1000 km the behavior is $\sim \delta^{2/3}$,  
corresponding to the Richardson's law; asymptotic
saturation sets in beyond this range (fully uncorrelated velocities).

\section{Discussion and conclusions}
\label{sec:conclusions} 

We have revisited the dispersion properties of the EOLE Lagrangian
dataset using a new approach, using Finite-Scale Lyapunov Exponent,
that is better suited to analyze scale dependent properties than
standard tools that were used, e.g., by ML in a previous study of the
same dataset.  We were motivated by the fact that ML found results
supporting a $k^{-3}$ inertial range between 100 and 1000~km, whereas
more recent studies based on aircraft data found a $k^{-5/3}$ behavior
in the same range of scales.

Our main result of our improved analysis is that the EOLE dataset 
supports a $k^{-5/3}$ behavior in the range 100-1000~km as shown
by the scaling properties of FSLE in this range indicating
Richardson's superdiffusion. At distances smaller than 100~km, our
results suggest an exponential separation with an e-folding time 
of about one day, in rough agreement with ML. At scales larger 
1000~km, the dispersion tends to a standard diffusion before
saturating at the planetary scale. Since the large-scale flow is 
dominated by the meandering zonal circulation, estimated diffusion 
coefficient is 10 times larger for total dispersion ($10^7$~\mms)
than for meridional dispersion ($10^6$~\mms).

Our result is compatible with an inverse 2D energy cascade in the range
100-1000~km or with the recently proposed alternative of a direct
energy cascade (Lindborg and Cho, 2000).  Our study of the EOLE
experiment has shown that this still unparalleled dataset of Lagrangian
trajectories in the atmosphere is in agreement with results obtained
from aircraft data.  The challenge is now to compare these trajectories
with the global wind fields produced by the recent reanalysis by
operational weather centers.

\bigskip

\begin{acknowledgments} 
We warmly thank G. Boffetta, F. D'Andrea, V. Daniel, B. Joseph, A. Mazzino, 
F. Vial for interesting discussions and suggestions. 
G.L. thanks the European Science Foundation for financial support through 
a TAO Exchange Grant 1999, and the LMD-ENS (Paris) for hosting him. 
\end{acknowledgments}

\newpage


\centerline{FIGURE CAPTIONS}

\bigskip
Figure 1. Mean square balloon separation.
 The four curves refers to 4 different initial thresholds: 
 25 km (a), 50 km (b), 100 km (c) and 200 km (d).
 All the curves but for 25 km have been shifted in time in order 
 to collapse together for $\langle R^2 \rangle > 10^7$~km$^2$.  
 The eddy diffusion coefficient corresponding to the indicated slope
 is  $D_E \simeq 2.9 \; 10^6$~\mms. Units: time in days, $R^2$ in km$^2$.

\bigskip
Figure 2. Mean logarithmic growth of the balloon separation ($+$). 
 The initial separation is $\le$ 25 km. 
 The two clusters 'A' ($\ast$) and 'B' ($\times$) have 4
 balloons each and were launched with a 3-day lag in November 1971. 
 The $e$-folding time of the exponential growth 
 is $\simeq 0.4$ day. The two horizontal lines mark the range 
 100-1000 km (natural logarithm units). The straight line ML is the result 
 found by Morel and Larchv\^eque (1974) in their fig. 8. 
Units: time in days, $R$ in km.

\begin{figure}[hbt]
\includegraphics[angle=0,width=17cm,clip]{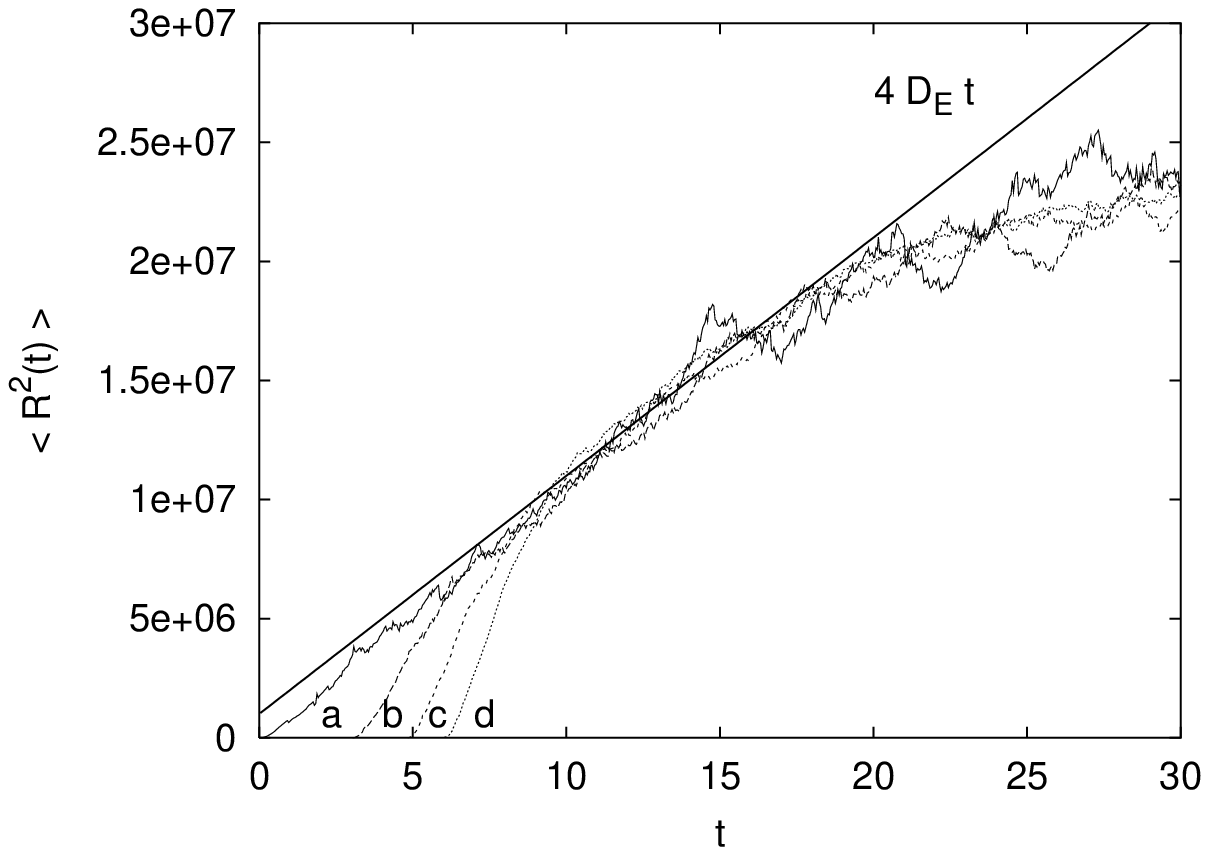}
\caption{
}
\label{fig:eoledisp100}
\end{figure}

\begin{figure}[hbt]
\includegraphics[angle=0,width=17cm,clip]{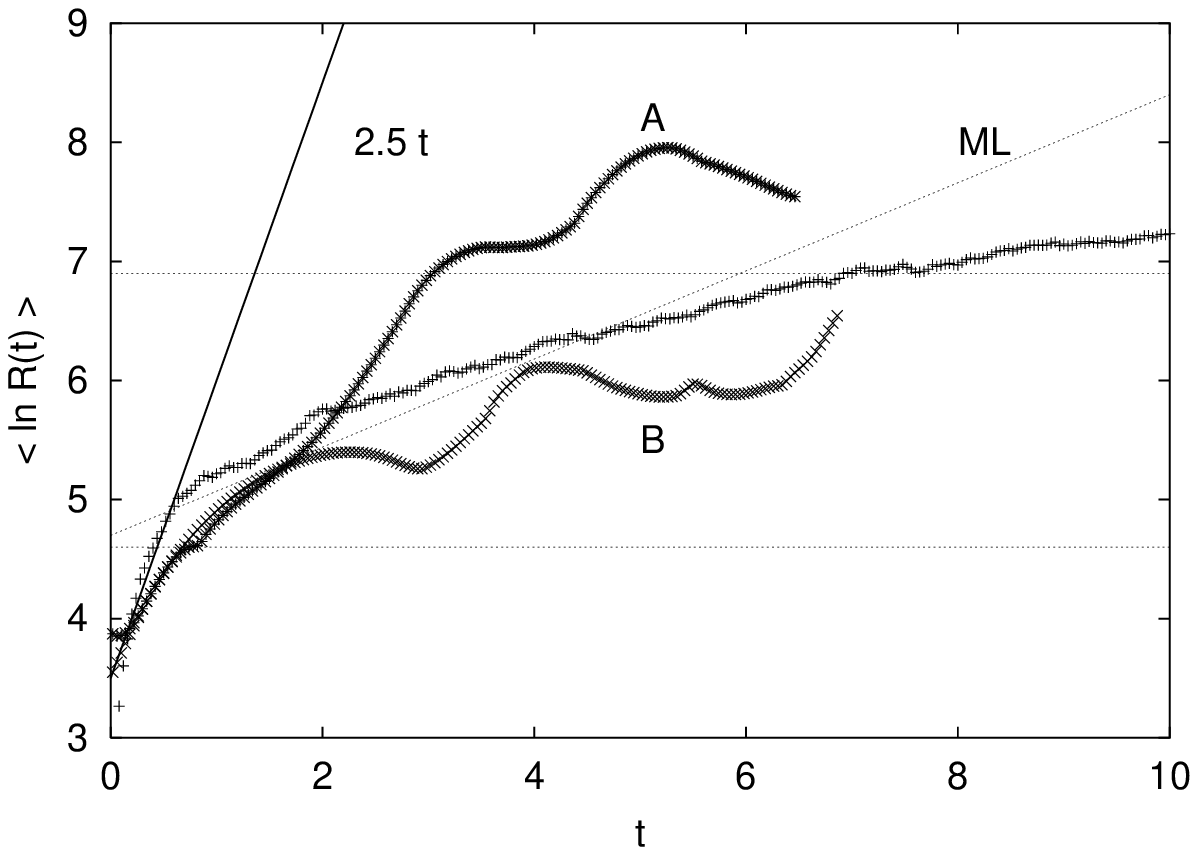}
\caption{
}
\label{fig:lyap}
\end{figure}

\begin{figure}[hbt]
\includegraphics[width=17cm]{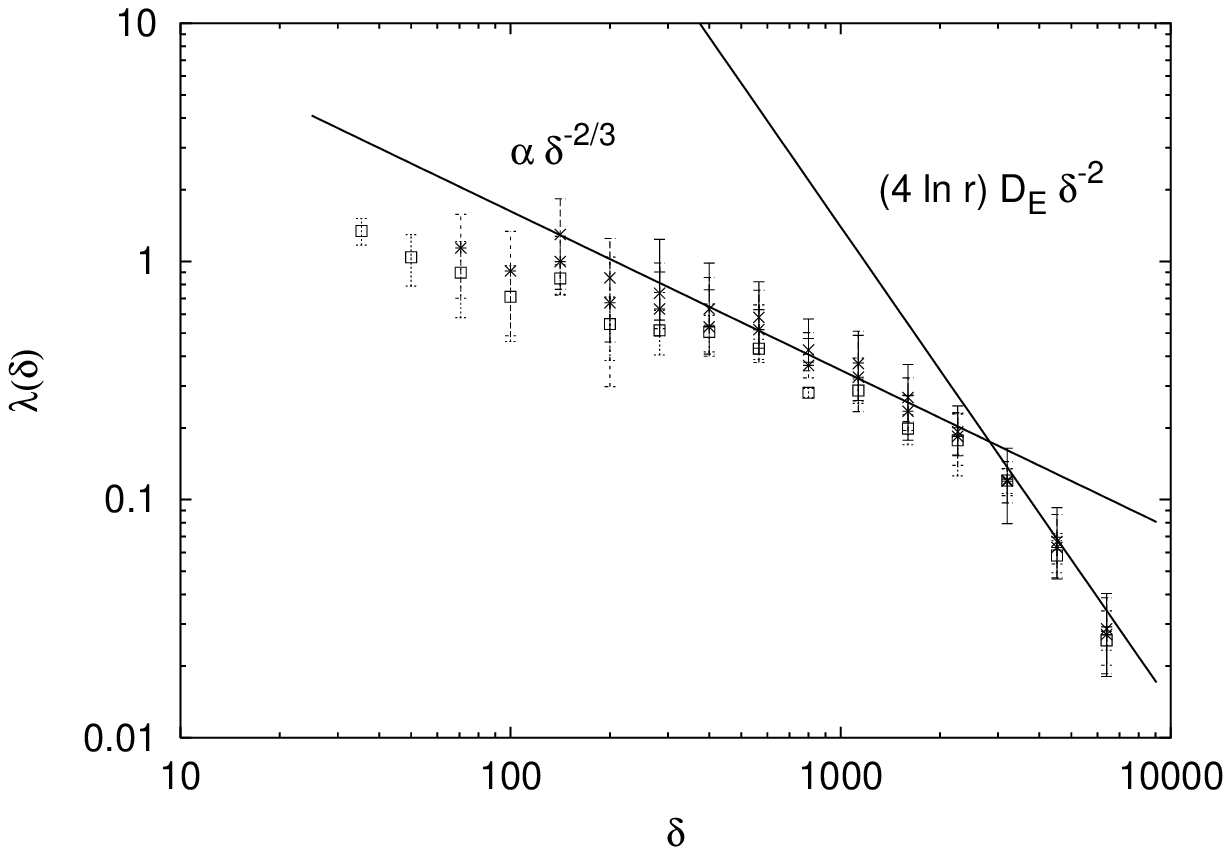}
\caption{
} 
\label{fig:eolefsle3}
\end{figure}

\begin{figure}[hbt]
\includegraphics[width=17cm]{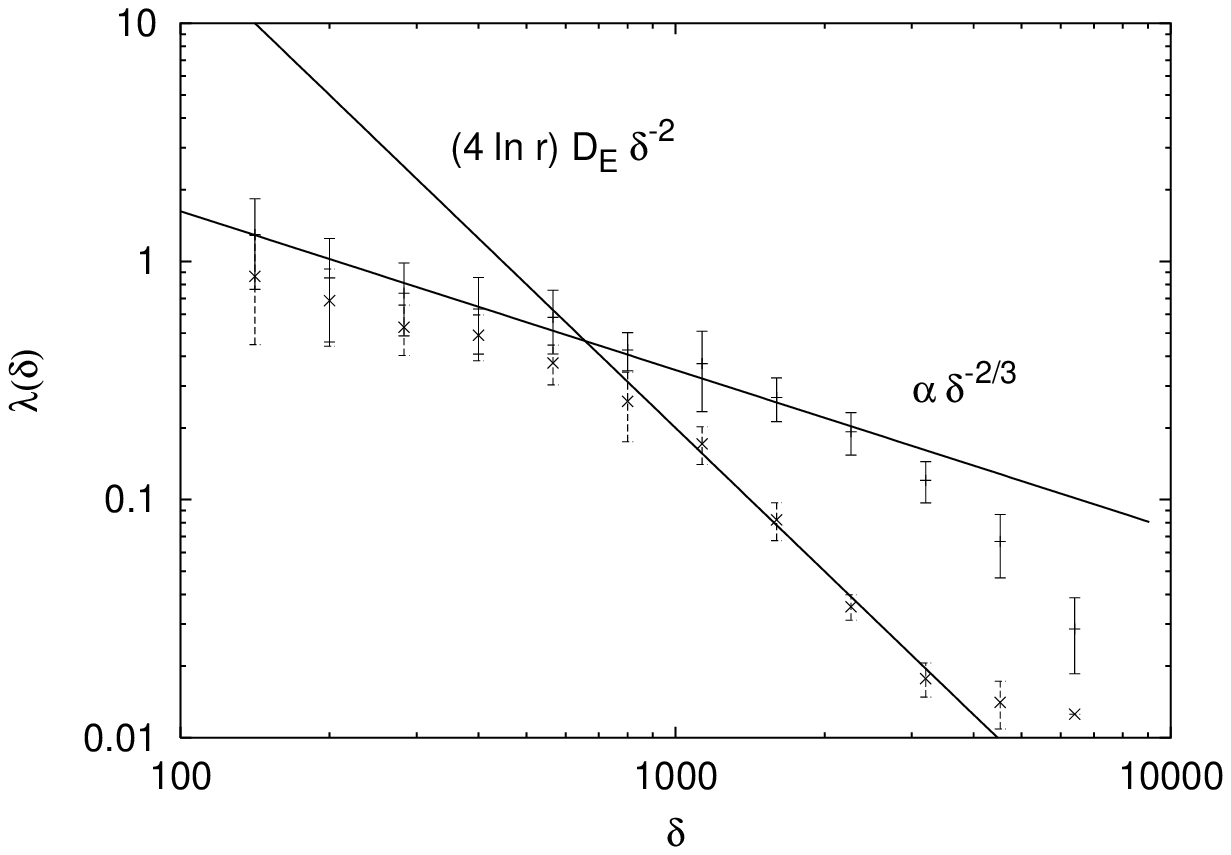}
\caption{
}
\label{fig:eolefsle100}
\end{figure}

\begin{figure}[hbt]
\includegraphics[width=17cm,clip]{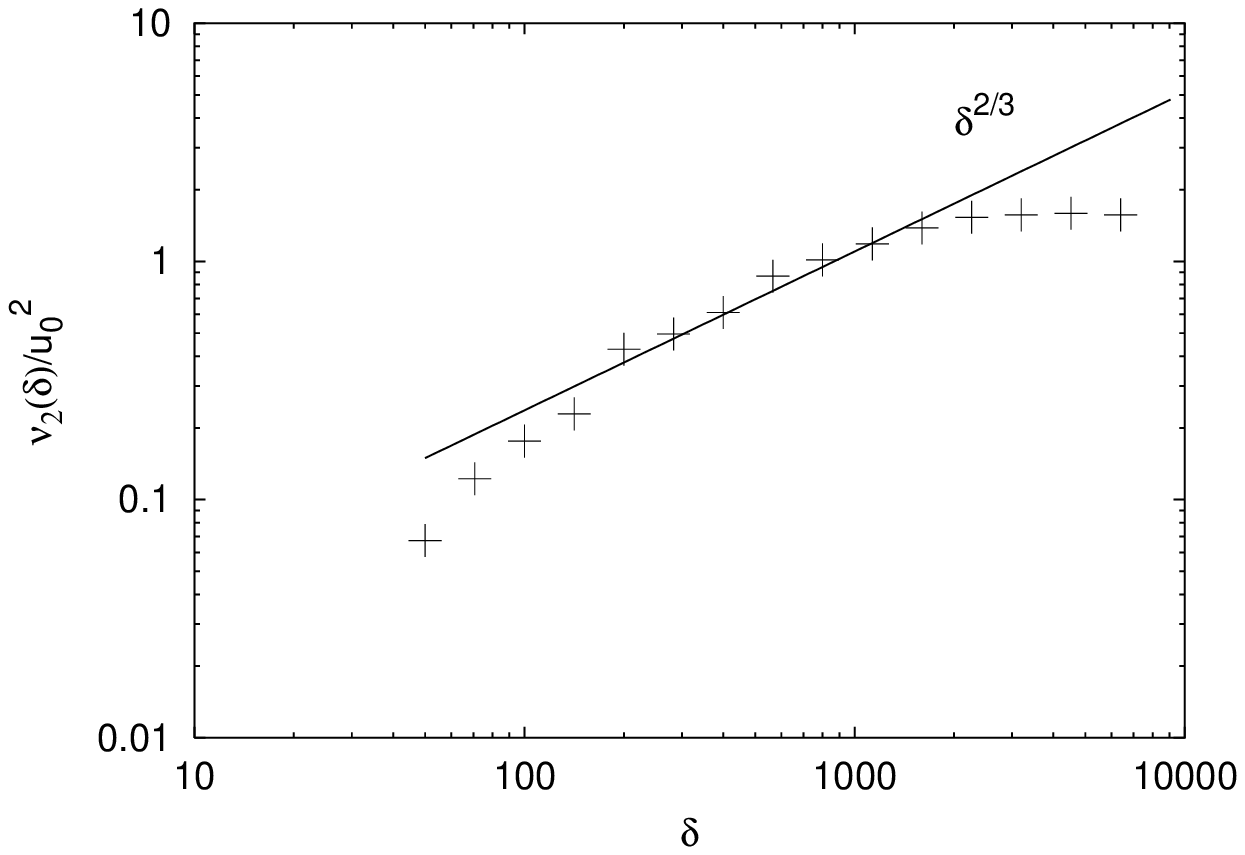}
\caption{
} 
\label{fig:rv}
\end{figure}

\bigskip
Figure 3. FSLE of the balloon pairs, four curves 
with the same four initial thresholds as in Figure \ref{fig:eoledisp100}. 
The eddy diffusion coefficient is $D_E \simeq 10^7$ \mms. The quantity  
$\alpha^3$ gives the order of magnitude of the relative kinetic energy growth 
rate (for unit mass) between balloons in the Richardson's regime 
$\sim \delta^{-2/3}$. 
Units: $\delta$ in km, $\lambda$ in day$^{-1}$.

\bigskip
Figure 4. FSLE of the balloon pairs, describing total ($-$) and
meridional ($\times$) dispersion, with initial threshold $100$ km.  
The meridional FSLE is
$\lambda_{mer}$ defined in (\ref{eq:fslemer}). 
The meridional eddy diffusion
coefficient is $D_E \simeq 1.5 \, 10^6$ \mms. 
Units: $\delta$ in km, $\lambda$ in day$^{-1}$.

\bigskip
Figure 5. FSRV of the balloon pairs for initial threshold 50 km. 
The reference velocity is $u_0=$100~km/hour.  The slope $2/3$
corresponds to the Richardson's law. Units: $\delta$ in km.

\newpage


\begin{longtable}{|l|l|l|l|l|} 
\caption{Number of balloon pairs analyzed for each scale during the 
computation of the FSLE. The first column is the order of the scale 
$\delta_n = r^n \delta_0$, with $r=\sqrt{2}$; 
the second, third, fourth and fifth column refer 
to the initial thresholds $\delta_0$=25 km, 50 km, 100 km and 200 km, 
respectively.
}\\
\hline
 n & 25 km & 50 km & 100 km & 200 km \\ \hline
 0 & 495 & 1037 & 2025 & 3979 \\
 1 & 344 & 782  & 1670 & 3649 \\
 2 & 391 & 867  & 1806 & 3699 \\
 3  & 414 & 895  & 1855 & 3764 \\
 4  & 440 & 951  & 1877 & 3722 \\
 5  & 442 & 955  & 1892 & 3687 \\
 6  & 456 & 950  & 1857 & 3563 \\
 7  & 442 & 944  & 1829 & 3471 \\
 8  & 448 & 928  & 1749 & 3327 \\
 9  & 440 & 906  & 1703 & 3131 \\
 10 & 428 & 865  & 1617 & 2648 \\
 11 & 418 & 845  & 1511 &\\
 12 & 397 & 794  & 1277 &\\
 13 & 389 & 756 &&\\
 14 & 368 & 642 &&\\
 15 & 346 &&&\\
 16 & 290 &&&\\  \hline

\end{longtable}

\newpage


\end{document}